\documentclass[doublecol,linenumbers]{epl2} 

\usepackage{graphicx}
\usepackage{epstopdf}
\usepackage{textcomp}
\usepackage{lipsum}
\usepackage{dblfloatfix}
\usepackage{fixltx2e}
\usepackage{color}
\usepackage[square,comma,numbers,sort&compress]{natbib} 
\newcommand{\infnge}{2}
\newcommand{\hzdr}{1}
\newcommand{\tudresden}{6}
\newcommand{\infnpd}{3}
\newcommand{\unipd}{4}
\newcommand{\edinburgh}{5}

\newcommand{\lngs}{9}
\newcommand{\atomki}{8}
\newcommand{\infnto}{10}
\newcommand{\infnmi}{12}
\newcommand{\unimi}{11}
\newcommand{\roma}{13}
\newcommand{\teramo}{15}

\newcommand{\gssi}{7}
\newcommand{\infnba}{14}

\begin{document}
\title{Effect of beam energy straggling on resonant yield in thin gas targets: The cases $\mathbf{^{22}Ne(p,\gamma)^{23}Na}$ and $\mathbf{^{14}N(p,\gamma)^{15}O}$}
\shorttitle{Effect of beam energy straggling on resonant yield in thin gas targets}
\author{
	D.\,Bemmerer\inst{\hzdr}\thanks{email: d.bemmerer@hzdr.de} \and
	F.\,Cavanna\inst{\infnge} \and
	R.\,Depalo \inst{\infnpd,\unipd} \and
	M.\,Aliotta \inst{\edinburgh} \and
	M.\,Anders\inst{\hzdr,\tudresden} \and
	A.\,Boeltzig \inst{\gssi}
	C.\,Broggini \inst{\infnpd} \and
	C.\,Bruno \inst{\edinburgh} \and
	A.\,Caciolli \inst{\infnpd} \and
	P.\,Corvisiero \inst{\infnge} \and
	T.\,Davinson \inst{\edinburgh} \and
	Z.\,Elekes \inst{\atomki} \and
	F.\,Ferraro \inst{\infnge}
	A.\,Formicola \inst{\lngs} \and
	Zs.\,F\"ul\"op \inst{\atomki} \and
	G.\,Gervino \inst{\infnto} \and
	A.\,Guglielmetti \inst{\unimi,\infnmi} \and
	C.\,Gustavino \inst{\roma} \and
	Gy.\,Gy\"urky \inst{\atomki} \and
	R.\,Menegazzo \inst{\infnpd} \and
	V.\,Mossa \inst{\infnba} \and
	F.R.\,Pantaleo \inst{\infnba} \and
	P.\,Prati \inst{\infnge} \and
	D.A.\,Scott \inst{\edinburgh} \and
	O.\,Straniero \inst{\lngs,\teramo} \and
	T.\,Sz\"ucs \inst{\atomki} \and
	M.P.\,Tak\'{a}cs\inst{\hzdr,\tudresden} \and
	D.\,Trezzi \inst{\infnmi}
	(LUNA\,collaboration)
	}                     

\institute{
	\inst{1} Helmholtz-Zentrum Dresden-Rossendorf (HZDR), Dresden, Germany 
	\\ \inst{2} Dipartimento di Fisica, Universit\`a di Genova, and Istituto Nazionale di Fisica Nucleare (INFN), Sezione di Genova, Genova, Italy 
	\\ \inst{3} INFN Sezione di Padova, Padova, Italy 
	\\ \inst{4} Dipartimento di Fisica, Universit\`a di Padova, Padova, Italy 
	\\ \inst{5} SUPA, School of Physics and Astronomy, University of Edinburgh, Edinburgh, United Kingdom 
	\\ \inst{6} Technische Universit\"at Dresden, Dresden, Germany 
	\\ \inst{7} Gran Sasso Science Institute, L'Aquila, Italy 
	\\ \inst{8} Institute of Nuclear Research (ATOMKI), Debrecen, Hungary 
	\\ \inst{9} INFN, Laboratori Nazionali del Gran Sasso, Assergi, Italy 
	\\ \inst{10} Dipartimento di Fisica Sperimentale, Universit\`a di Torino, and INFN Sezione di Torino, Torino, Italy 
	\\ \inst{11} Universit\`a degli Studi di Milano, Milano, Italy 
	\\ \inst{12} INFN, Sezione di Milano, Milano, Italy 
	\\ \inst{13} INFN, Sezione di Roma "La Sapienza", Roma, Italy, 
	\\ \inst{14} Universit\`a degli Studi di Bari and INFN, Sezione di Bari, Bari, Italy 
	\\ \inst{15} Osservatorio Astronomico di Collurania, Teramo, Italy 
}
\shortauthor{D. Bemmerer \etal\ (LUNA collab.)}

\abstract{
When deriving resonance strengths using the thick-target yield approximation, for very narrow resonances it may be necessary to take beam energy straggling into account. This applies to gas targets of a few keV width, especially if there is some additional structure in target stoichiometry or detection efficiency. The correction for this effect is shown and tested on recent studies of narrow resonances in the ${\rm^{22}Ne(p,\gamma)^{23}Na}$ and ${\rm^{14}N(p,\gamma)^{15}O}$ reactions.
} 
\maketitle

\section{Introduction} In the thick target study of a radiative proton capture reaction, the yield on top of the resonance plateau $Y_{\rm max}$ is used to obtain the resonance strength $\omega\gamma$ as:
\begin{equation}
\label{eq:Thick}
\omega\gamma = \omega \frac{\Gamma_p \Gamma_\gamma}{\Gamma_p + \Gamma_\gamma} = \frac{2\,Y_{\rm max}\,\epsilon_{\rm R}}{\lambda^2_{\rm R}}\frac{m_{\rm t}}{m_{\rm t}+m_{\rm p}}
\end{equation}
where $\omega$ is the statistical factor, $\Gamma_{p,\gamma}$ are the proton and $\gamma$-ray widths of the resonance under study, 
$\epsilon_{\rm R}$ is the effective stopping power in the laboratory system, $\lambda^2_{\rm R}$ is the squared de Broglie wavelength at the center-of-mass resonance energy, and $m_{\rm t}$ and $m_{\rm p}$ are the masses of target and projectile, respectively. 
Equation (\ref{eq:Thick}) is applicable when the energetic target thickness $\Delta E$ is large compared to the total width $\Gamma$=$\Gamma_p$+$\Gamma_\gamma$. For intermediate cases, the strength from eq.~(\ref{eq:Thick}) must be multiplied with a factor $\frac{\pi}{2}/\arctan(\Delta E/\Gamma)$ that approaches unity for $\Delta E \gg \Gamma$ \cite{Iliadis15-Book}. 

\begin{figure}[tb]
\includegraphics[width=\columnwidth]{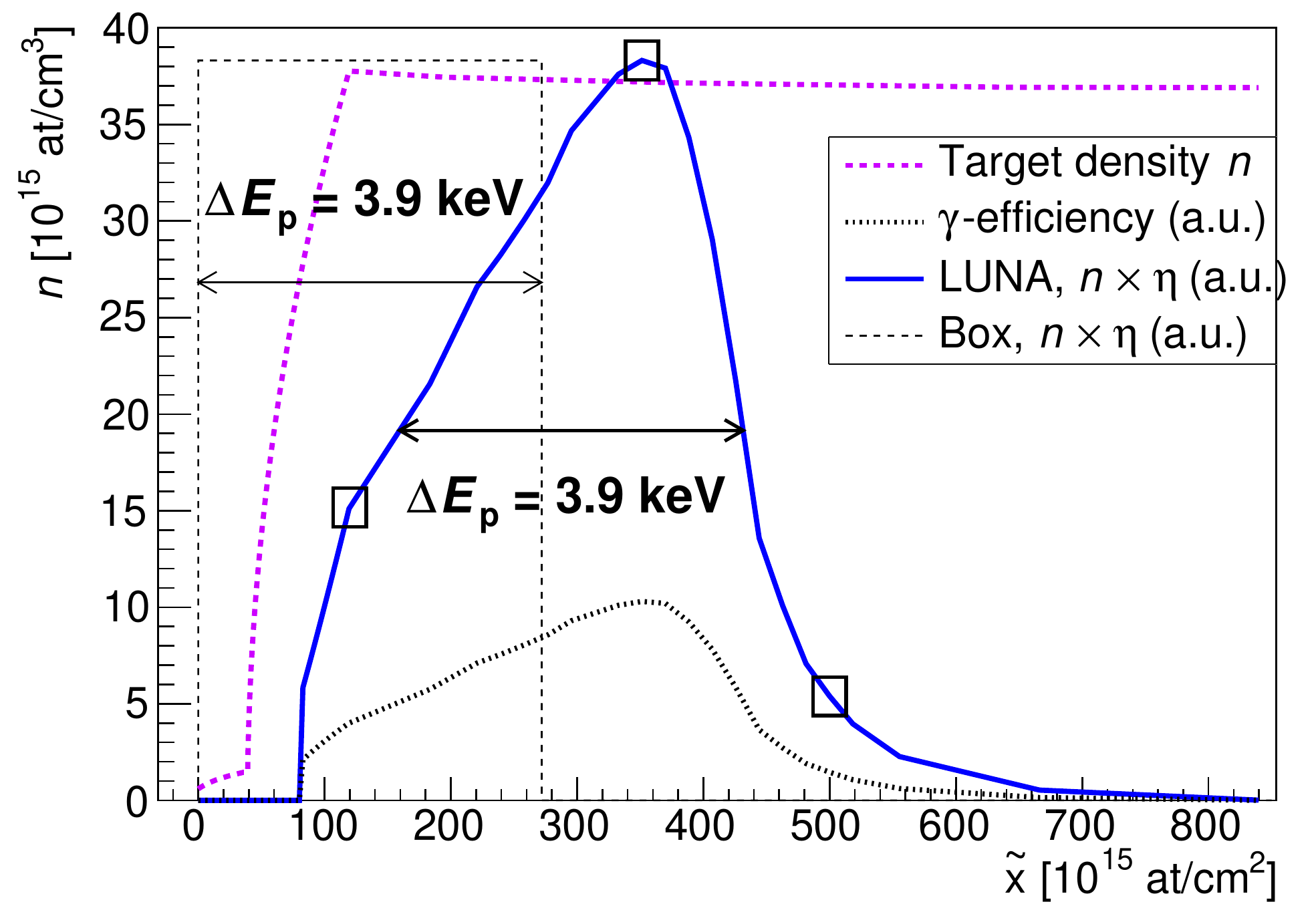}
\caption{LUNA gas target density ($n$, purple dashed line), $\gamma$-detection efficiency ($\eta$, black dotted line, for $E_\gamma$ = 478\,keV, detector Ge55 \cite{Cavanna15-PRL,Depalo16-PRC}), and product of the two ($n\times\eta$, blue straight line) as a function of position $\tilde{x}$ in the target. An ideal box shape of same  width and maximum  is also shown (black dashed line). See text for details.
}
\label{fig:Density}
\end{figure}

For the experiment of Refs.~\cite{Cavanna15-PRL,Depalo16-PRC}, $\Delta E \sim$ 3.9\,keV (Figure \ref{fig:Density}).
The three new resonances reported in Refs.~\cite{Cavanna15-PRL,Depalo16-PRC} 
correspond to excited states in $^{23}$Na at $E_x$ = 8944 ($J^\pi$=3/2$^+$), 8975 (5/2$^+$), and 9042 (7/2$^+$ or 9/2$^+$) keV, respectively (spin assignments from Ref.~\cite{Jenkins13-PRC}). They were not observed in a nuclear resonance fluorescence experiment with bremsstrahlung up to 10.4 MeV and a typical sensitivity of 0.1\,eV for the partial $\gamma$-width  \cite{Vodhanel84-PRC}. Assuming 1\% ground state branching, this leads to a limit of $\Gamma_\gamma<$10\,eV.
In a $^{22}$Ne($^3$He,d)$^{23}$Na experiment \cite{Hale01-PRC}, values or upper limits between 2.3$\times$10$^{-9}$\,eV and 1.1$\times$10$^{-6}$\,eV are reported for the proton widths of these states. 

As a result, $\Delta E \gg \Gamma_\gamma+\Gamma_p=\Gamma$ is found for all three resonances, and they seem to be textbook \cite{Iliadis15-Book} cases for the applicability of the thick-target yield formula eq.~(\ref{eq:Thick}). This is the assumption made explicitly in Refs. \cite{Cavanna15-PRL,Depalo16-PRC}.
However, this is inappropriate, as will be shown below.

\begin{figure}[tb]
\includegraphics[width=\columnwidth]{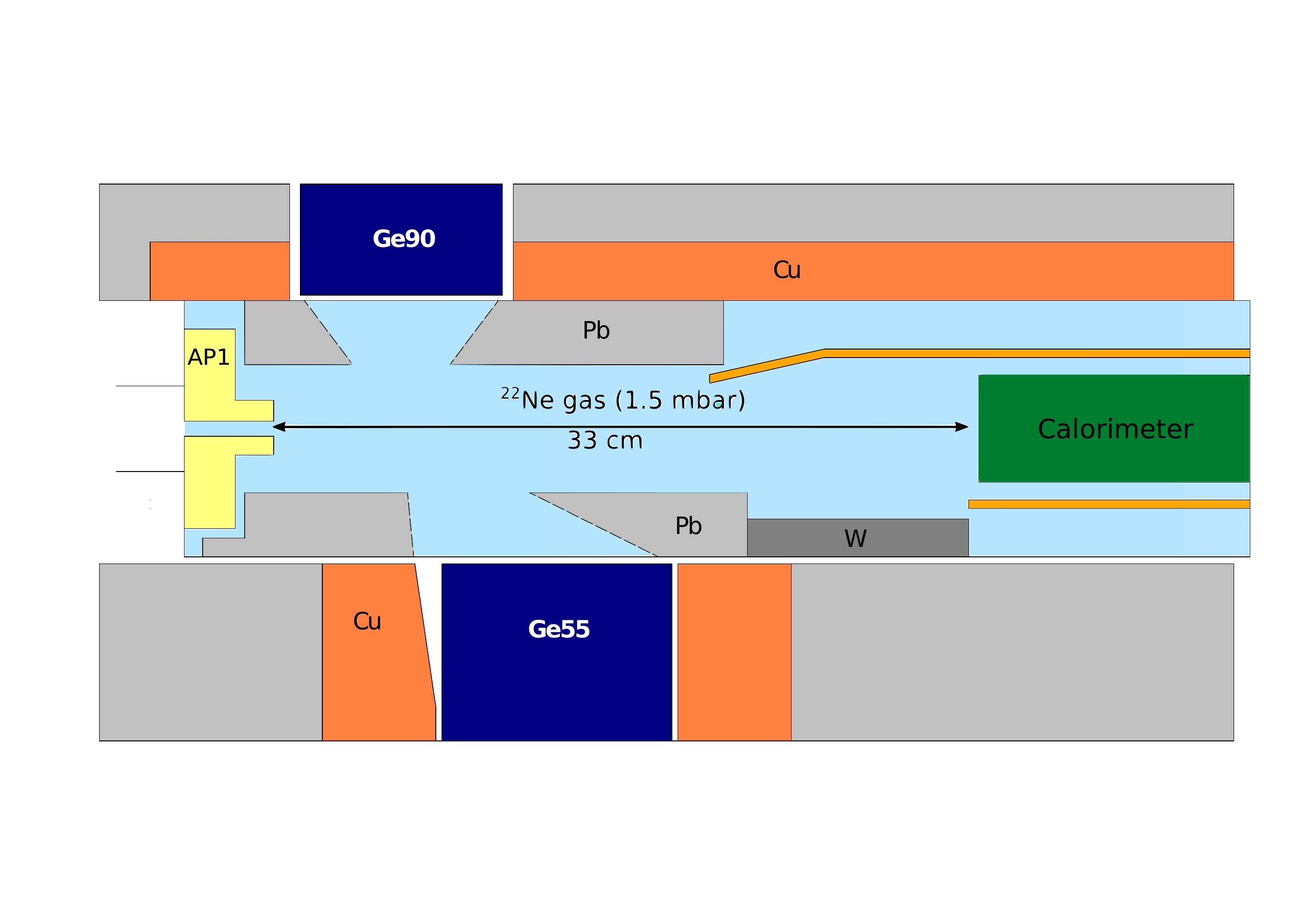}
\caption{Sketch of the experimental setup. The ion beam enters from the left and is stopped on the beam calorimeter. The color encodes the HPGe detectors (Ge55 and Ge90, dark blue), the copper (Cu, orange), lead (Pb, light grey), and tungsten (W, dark grey) shielding, the gas limiting aperture (AP1, yellow), the beam calorimeter (green), and the $^{22}$Ne gas (light blue).
}
\label{fig:Setup}
\end{figure}

\section{Derivation of the correction} The experimental setup (Figure \ref{fig:Setup}) is characterized by a windowless, static-type $^{22}$Ne gas target with two HPGe detectors, collimated to give effective detection angles of 55$^\circ$, and 90$^\circ$, respectively.

The profiles of effective target density $n$ and $\gamma$-ray detection efficiency $\eta$ for the present case \cite{Cavanna14-EPJA,Cavanna15-PRL,Depalo16-PRC,Cavanna15-PRL-Erratum} are shown in Figure \ref{fig:Density}. For their product $n \times \eta$, a full width at half maximum of $\Delta E_p$ = 3.9 keV is found. There is significant structure in the $\gamma$-ray detection efficiency curve as a consequence of the collimation. For the sake of the discussion, an ideal box profile of the same width and with a height corresponding to the maximum of the LUNA $n\times\eta$ profile has been added (Figure \ref{fig:Density}). This profile starts at the same place as the real gas target, i.e. in the connection tube between final collimator and first pumping stage. 

\begin{figure}[tb]
\includegraphics[width=\columnwidth]{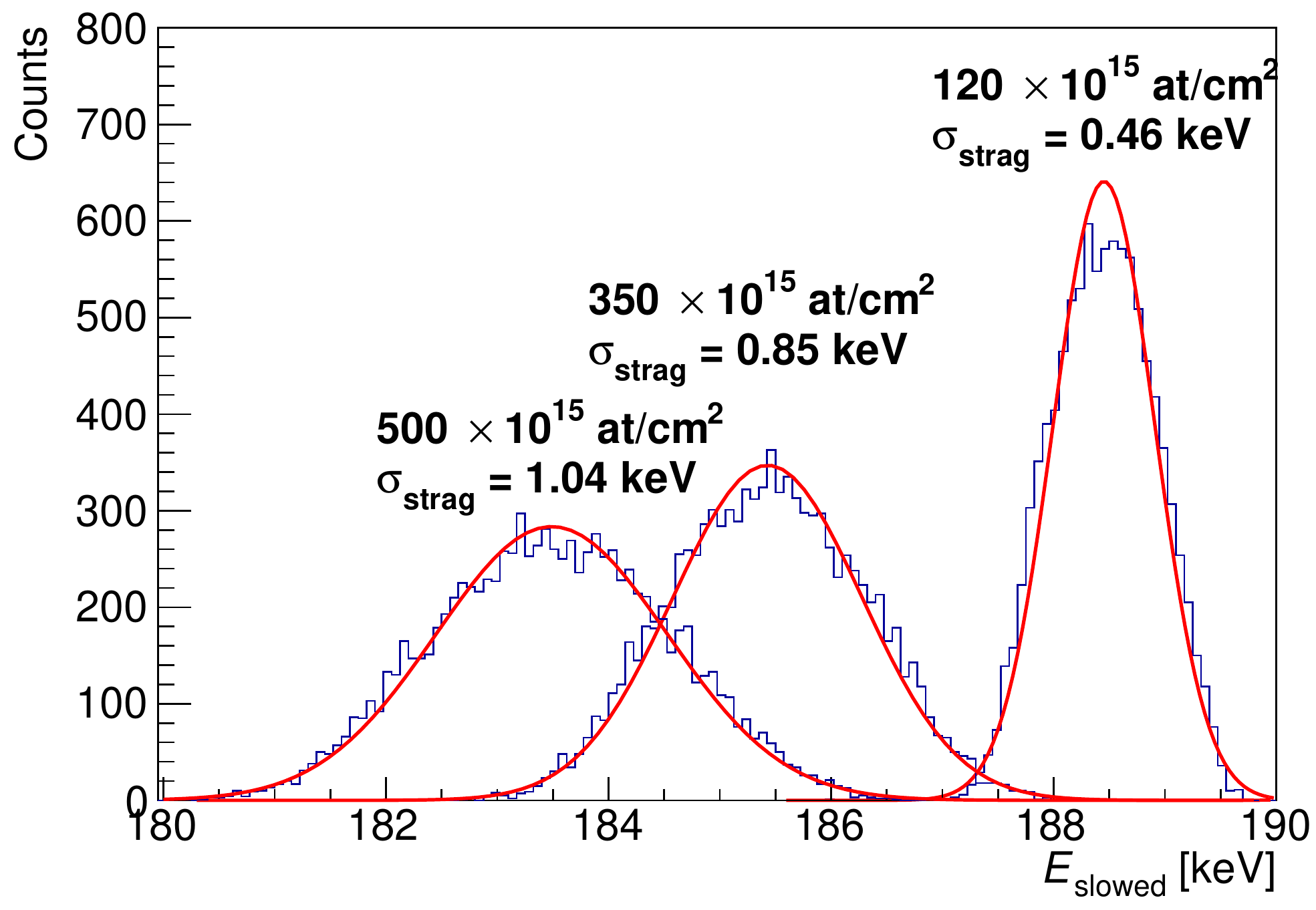}
\caption{Energy straggling of the slowed proton beam, simulated by SRIM \cite{Ziegler10-NIMB}, at the three points marked in Figure \ref{fig:Density}.}
\label{fig:SRIM}
\end{figure}

As a next step, the energy loss of an $E_p$ = 190 keV proton beam when passing through the $^{22}$Ne gas target is simulated by SRIM \cite{Ziegler10-NIMB}. The energy distribution of the slowed beam is almost Gaussian  (Figure \ref{fig:SRIM}) with mean energy $E_{\rm slowed}^{\rm mean}$ and straggling width $\sigma_{\rm strag}$, and an empirical parameterization for $\sigma_{\rm strag}(E_{\rm slowed}^{\rm mean})$ is derived. 

Using this information and in addition the energy spread of the proton beam from the accelerator (at LUNA, \linebreak $\sigma_{\rm beam}\le$0.1/2.355 keV \cite{Formicola03-NIMA}, much lower than the straggling width), an approximate energy distribution of the slowed beam at position $\tilde{x}$ in the target can be derived as:
\begin{equation}
\label{eq:Beamwidth}
f_{\rm beam}(E,E_{\rm slowed}^{\rm mean}) = \exp\left[ -\frac{(E-E_{\rm slowed}^{\rm mean})^2}{2\sigma^2_{\rm strag}(E_{\rm slowed}^{\rm mean})+2\sigma_{\rm beam}^2} \right]
\end{equation}

\begin{figure}[tb]
\includegraphics[width=\columnwidth]{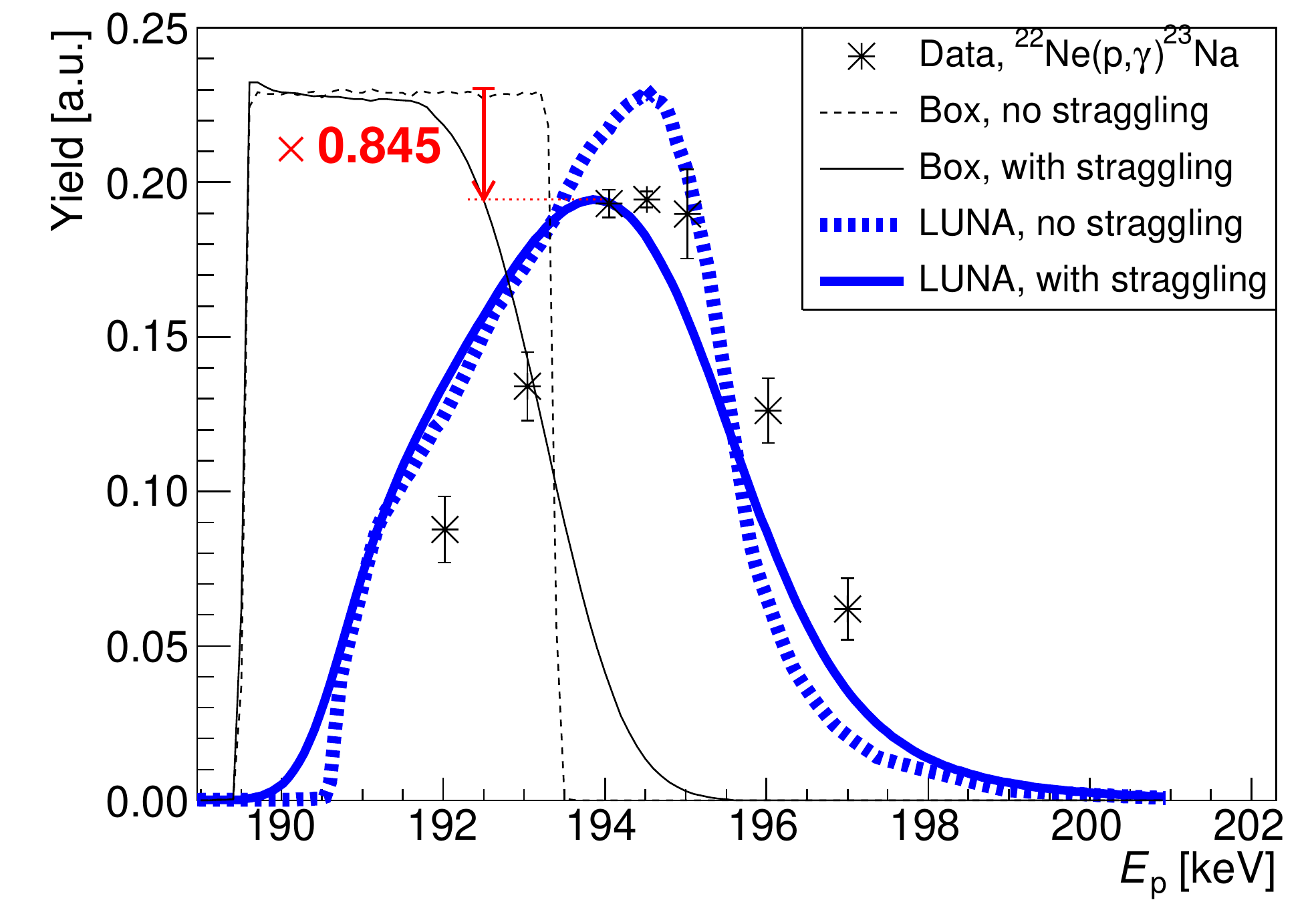}
\caption{Experimental yield (data points) of the $E_\gamma$ = 440\,keV transition in the 189.5\,keV $^{22}$Ne(p,$\gamma$)$^{23}$Na resonance, 
 compared with calculated yield curves without straggling (blue dotted line) and with straggling (blue full line). The experimental yield is about 15\% lower than expected for an ideal flat-top profile without straggling (thin dashed black line), or with straggling (thin black line). See text for details. }
\label{fig:Yield}
\end{figure}

Finally, the yield $Y(E_p)$ for a resonance scan performed on this target with this beam can be computed for each incident beam energy $E_p$, by numerically integrating the product of the  usual Breit-Wigner resonance shape $\sigma_{\rm BW}(E)$, $f_{\rm beam}(E,E_{\rm slowed}^{\rm mean})$, and the target and efficiency profiles $n(\tilde{x}) \times \eta(\tilde{x})$:
\begin{eqnarray}
Y(E_p) & = & \int\limits_{\tilde{x}=0}^{\tilde{x}=\tilde{x}_{\rm max}} d\tilde{x} \int\limits_{E=E_p}^{E=0} dE \nonumber \\[2mm] 
 & & \sigma_{\rm BW}(E) \; f_{\rm beam}(E,E_{\rm slowed}^{\rm mean}(\tilde{x})) \; n(\tilde{x}) \; \eta(\tilde{x})  
\end{eqnarray}
As a first step, this formula is applied to the ideal box profile for the new $E_p$ = 189.5\,keV resonance, first by artificially imposing $\sigma_{\rm strag} \equiv 0$ so that only the small beam energy spread remains (black thin dashed curve in Figure \ref{fig:Yield}), then including $\sigma_{\rm strag}$ from SRIM (black thin straight curve in Figure \ref{fig:Yield}). It is thus confirmed that in the textbook case, straggling does not significantly affect the plateau yield \cite{Iliadis15-Book}.

However, the picture changes when the realistic profile (blue curve in Figure \ref{fig:Density}) is used: Here, the artificial case without straggling (thick blue dashed curve in Figure \ref{fig:Yield}) is significantly higher than the realistic case with straggling (thick blue full curve in Figure \ref{fig:Yield}). 

As a consequence, the resonance strength data by Refs.~\cite{Cavanna15-PRL,Depalo16-PRC} must be corrected upwards, dividing by the factor $C$ that reflects the reduction in the yield profile observed when correctly including the energy straggling of the proton beam. Repeating the calculation also for the other two resonances, new values for the resonance strength are found (Table \ref{Table:omegagamma}). Since this correction is a purely calculated one, a conservative error bar of 50\% is assigned to the correction. 

The present, corrected resonance strengths are in agreement with independent, new data on the resonances at 156.2 and 189.5 keV that have been reported in the mean time by the TUNL group \cite{Kelly17-PRC}. The upper limits reported in Refs.~\cite{Cavanna15-PRL,Depalo16-PRC} are not updated here, because much more restrictive upper limits may be expected from an experiment on the same reaction with a new setup \cite{Ferraro18-EPJA}.

\begin{table*}[tb]
\caption{Original  \cite{Cavanna15-PRL,Depalo16-PRC} and corrected \cite{Cavanna15-PRL-Erratum} values for the resonance strength $\omega\gamma$, and straggling correction factor $C$. For the corrected values, the statistical error bars (unchanged) and the systematical error bars (including the new correction) are given separately.}
\center
\begin{tabular}{|c|c|c|c|r|r|}
\hline
$E_p$ [keV] & $\omega\gamma_{\rm orig}$ [eV] \cite{Cavanna15-PRL,Depalo16-PRC} & $C$ & $\omega\gamma_{\rm corr} $ [eV] \cite{Cavanna15-PRL-Erratum} & stat. & syst. \\ \hline
156.2 & (1.48$\pm$0.10) $\times$10$^{-7}$ & 0.845 & $1.8  \times$10$^{-7}$ & 6\% & 8\%\\
189.5  & (1.87$\pm$0.06)$\times$10$^{-6}$ & 0.850 & $2.2 \times$10$^{-6}$ & 2\% & 8\%  \\
259.7 & (6.89$\pm$0.16)$\times$10$^{-6}$ & 0.841 & $8.2 \times$10$^{-6}$ & 1\% & 8\% \\

\hline
\end{tabular}
\label{Table:omegagamma}
\end{table*}

\section{Application to the $^{14}$N(p,$\gamma$)$^{15}$O reaction} 
In order to gain a further cross-check of the validity of the present correction, a run of this measurement campaign \cite{Cavanna15-PRL,Depalo16-PRC} is re-analyzed here. During this run aimed at extending the $\gamma$-detection efficiency curve to higher energies, the  gas target was filled with 3.5\,mbar nitrogen, and the $E_p$ = 278\,keV resonance in the $^{14}$N(p,$\gamma$)$^{15}$O reaction was excited. It is well known that the $\gamma$ rays de-exciting this 1/2$^+$ resonance are isotropic \cite{Ajzenberg13_15_91-NPA,Costantini03-Diss}, and the branching for the strongest branch, consisting of two $\gamma$ rays at 1384 and 6172\,keV, respectively, has recently been remeasured very precisely, to (58.3$\pm$0.3)\% \cite{Marta08-PRC,Marta11-PRC,Daigle16-PRC}.

\begin{figure}[tb]
\includegraphics[width=\columnwidth]{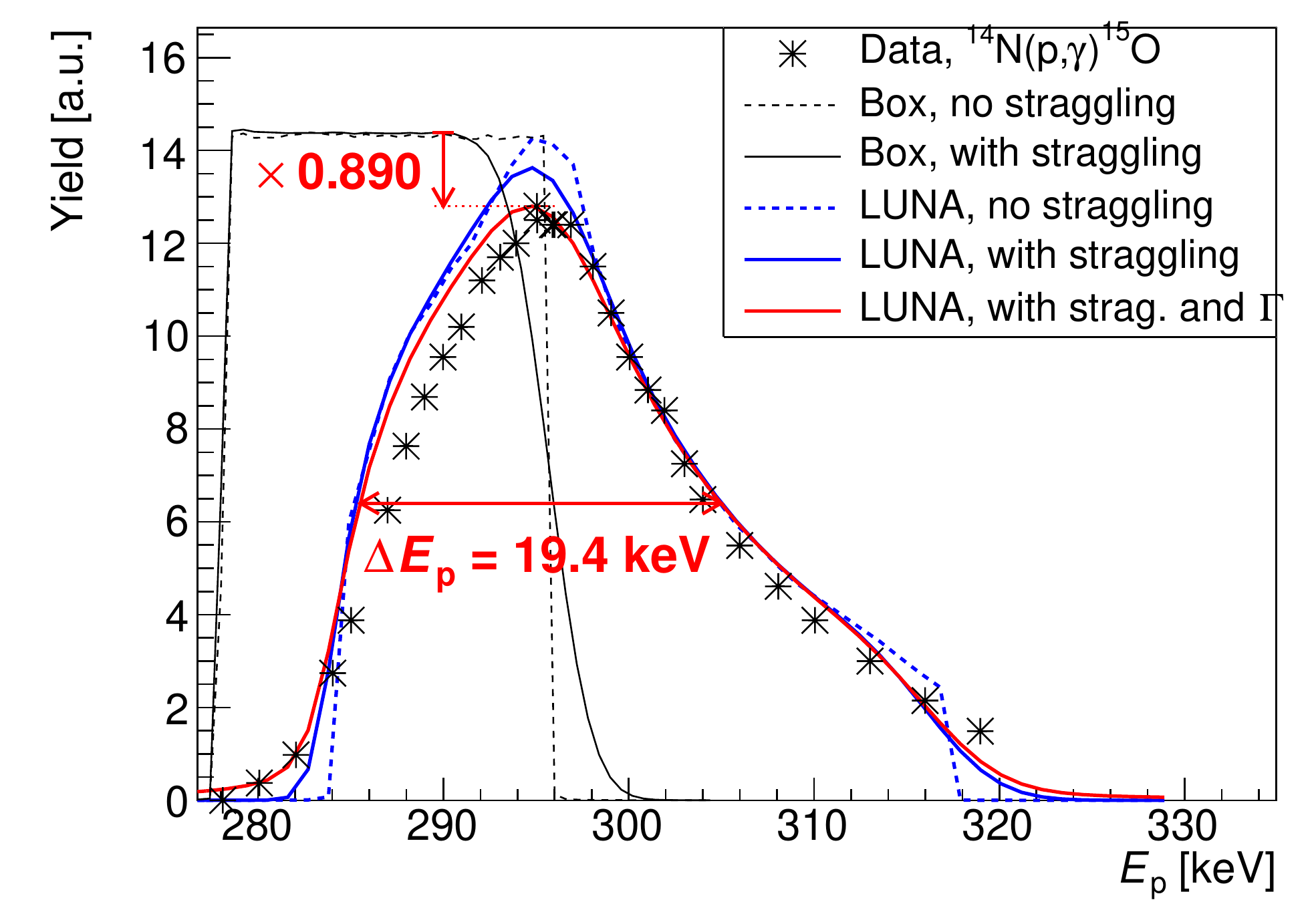}
\caption{Same as Figure \ref{fig:Yield}, but for the 278 keV resonance in the $^{14}$N(p,$\gamma$)$^{15}$O reaction. The final correction includes not only the straggling, but also a total resonance width of 1.07\,keV in the laboratory system.}
\label{fig:14N}
\end{figure}

This resonance has a non-negligible total width, $\Gamma_p$ = (0.999$\pm$0.046) keV \cite{Daigle16-PRC} that must be included in eq.~(\ref{eq:Beamwidth}). Even though the target is much wider than the resonance, $\Delta E_p\sim$19.4\,keV, again due to the structure of the efficiency curve, a correction is necessary, by dividing by 0.890. Using the known branching \cite{Marta08-PRC,Marta11-PRC,Daigle16-PRC}, a resonance strength of (12.7$\pm$0.3$_{\rm stat}$$\pm$1.0$_{\rm syst}$) meV is found. The systematic error includes 4\% for the $\gamma$-detection efficiency, 1\% for the beam intensity, 2.9\% for the stopping of protons in nitrogen \cite{Ziegler10-NIMB}, 0.5\% for the branching ratio, and 6\% for the present correction, in total 8\%.

The present new strength for the 278 keV resonance in the $^{14}$N(p,$\gamma$)$^{15}$O reaction is consistent with a recent very precise re-evaluation, which resulted in a value of (12.6$\pm$0.3) meV \cite{Daigle16-PRC}. 

\section{Discussion} 

The presently derived correction for narrow resonance yields, while general in nature, is expected to play a significant role only for the case of targets that are extended over such a large space that inhomogeneities in both target thickness and detection efficiency must be considered. In practice this will mainly apply to windowless, static-type gas targets, whose density-efficiency profile, $n(x) \times \eta(x)$, significantly deviates from an ideal box shape. 

Of all the LUNA experiments, this correction has a significant effect only for the HPGe-detector based phase of the study of the $\rm^{22}Ne(p,\gamma)^{23}Na$ reaction \cite{Cavanna15-PRL,Depalo16-PRC,Cavanna15-PRL-Erratum}.  In the other LUNA experiments \cite{Broggini18-PPNP}, either non-resonant processes were studied \cite[e.g.]{Formicola04-PLB,Lemut06-PLB,Bemmerer06-PRL,Caciolli11-AA,Anders14-PRL} or solid targets (where the detection efficiency inside the target is approximately constant and structure usually mitigated) were used \cite[e.g.]{Strieder12-PLB,Scott12-PRL,Bruno16-PRL}.

It should be noted that in these cases, the alternative approach to integrate the entire resonance scan \cite{Iliadis15-Book} cannot be applied due to the non-constant detection efficiency.

A static-type gas target should be designed in such a way that not only the total width of the resonance, but also the beam energy spread (loss and straggling) remain much smaller than any $n(\tilde{x})\times\eta(\tilde{x})$ structure. This is actually the case in forthcoming LUNA work on the ${\rm^{22}Ne(p,\gamma)^{23}Na}$ reaction with higher target density and very flat efficiency profile, due to the use of a $\gamma$-calorimeter \cite{Ferraro18-EPJA}.

\acknowledgments

Financial support by INFN, DFG (BE 4100-4/1), NAVI (HGF VH-VI-417), NKFIH (K120666), and DAAD fellowships at HZDR for F.C. and R.D. are gratefully acknowledged. 
%



\end{document}